\documentclass[twocolumn]{aastex61}

\definecolor{maroon}{rgb}{0.5, 0.0, 0.0}
\definecolor{dark-green}{rgb}{0.05, 0.5, 0.06}
\definecolor{dark-blue}{rgb}{0.0, 0.0, 0.5}
\definecolor{darkgray}{rgb}{0.85, 0.85, 0.85}

\usepackage{xcolor}
\usepackage{amsmath}

\received{\today}
\revised{\today}
\accepted{\today}

\submitjournal{ApJ}

\shorttitle{3-phase evolution}
\shortauthors{Heinemann et al.}

\begin{document}

\title{3-Phase Evolution of a Coronal Hole, Part I: 360$\arcdeg$ remote sensing and in-situ observations}

\correspondingauthor{Stephan G. Heinemann}
\email{stephan.heinemann@hmail.at}

\author{Stephan G. Heinemann}
\affil{University of Graz, Institute of Physics, Universit\"atsplatz 5, 8010 Graz, Austria }

\author{Manuela Temmer}
\affil{University of Graz, Institute of Physics, Universit\"atsplatz 5, 8010 Graz, Austria }

\author{Stefan J. Hofmeister}
\affil{University of Graz, Institute of Physics, Universit\"atsplatz 5, 8010 Graz, Austria }

\author{Astrid M. Veronig}
\affil{University of Graz, Institute of Physics, Universit\"atsplatz 5, 8010 Graz, Austria }


\author{Susanne Vennerstrom}
\affil{National Space Institute, DTU Space, Denmark }

\begin{abstract}
We investigate the evolution of a well-observed, long-lived, low-latitude coronal hole (CH) over 10 solar rotations in the year 2012. By combining EUV imagery from STEREO-A/B and SDO we are able to track and study the entire evolution of the CH having a continuous 360$\arcdeg$ coverage of the Sun. The remote sensing data are investigated together with in-situ solar wind plasma and magnetic field measurements from STEREO-A/B, ACE and WIND. From this we obtain how different evolutionary states of the CH as observed in the solar atmosphere (changes in EUV intensity and area) affect the properties of the associated high-speed stream measured at $1$AU. Most distinctly pronounced for the CH area, three development phases are derived: a) growing, b) maximum, and c) decaying phase. During these phases the CH area a) increases over a duration of around three months from about $1 \cdot 10^{10} \mathrm{km}^{2}$ to $6 \cdot 10^{10} \mathrm{km}^{2}$, b) keeps a rather constant area for about one month of $> 9 \cdot 10^{10} \mathrm{km}^{2}$, and c) finally decreases in the following three months below $1 \cdot 10^{10} \mathrm{km}^{2}$ until the CH cannot be identified anymore. The three phases manifest themselves also in the EUV intensity and in in-situ measured solar wind proton bulk velocity. Interestingly, the three phases are related to a different range in solar wind speed variations and we find for the growing phase a range of $460-600$~km~s$^{-1}$, for the maximum phase $600-720$~km~s$^{-1}$, and for the decaying phase a more irregular behavior connected to slow and fast solar wind speed of $350-550$~km~s$^{-1}$.

\end{abstract}


\section{Introduction} \label{sec:intro}


Coronal holes (CHs) are large-scale structures in the solar corona with one dominant magnetic polarity due to magnetic field lines that do not close within the CH nor within its vicinity. Along those open field lines, plasma can easily escape and radially flow outwards at high speeds \citep{wilcox68}. The fast plasma outflow from those localized coronal areas is related to an increased solar speed in interplanetary space and referred to as high-speed solar wind streams (HSS). Due to the interaction with the ambient slow solar wind, so-called stream interaction regions are formed. As CHs are long-lived features with lifetimes longer than several solar rotations, these stream interaction regions become corotating interaction regions. At a distance range of 1~AU, in-situ measurements of the solar wind plasma and magnetic field typically show that the CIR is dominated by a strong bump in the density and magnetic field, followed by a gradual increase in velocity that peaks several days after \citep{gosling96}. The production of shocks, compression and rarefaction regions associated to CIRs are well known sources of recurrent geomagnetic effects at Earth \citep[see e.g.,][]{alves06,kilpua17}. Especially during the late declining phase of the solar cycle, CIRs may cause weak to moderate geomagnetic storms \citep[e.g.,][]{verbanac11}. 

Besides their geomagnetic effects, CIRs strongly structure the interplanetary solar wind flow which has important consequences for the propagation behavior of coronal mass ejections (CMEs) and is one of the key parameters for Space Weather forecasting models. Current solar wind models still have uncertainties in simulating the background solar wind \citep[see e.g.,][]{gressl14,jian15}. Improving our knowledge about the sources of high-speed solar wind streams, i.e.\,the evolution of CHs, is of crucial importance to advance Space Weather forecasting.

CH structures are of low plasma density, and therefore of reduced emission, which makes them observable as dark regions in the extreme ultraviolet (EUV) and X$-$ray wavelength range \citep[e.g.,][]{schwenn06}. The identification and extraction is usually based on intensity-threshold methods \citep{2009krista, 2012rotter, 2014reiss, 2017hofmeister}. The observed CH areas are found to be well correlated with the solar wind speed measured in-situ at 1~AU \citep{1973krieger, 1976nolte, 2007vrsnak, 2012rotter, 2018hofmeister}. Changes in the photospheric magnetic field underlying a CH have consequences in the coronal expansion, hence, in its observed area and intensity, that shapes the solar wind outflow and its signatures at 1~AU \citep{wang90,gosling96}. However, as observations are usually limited to single point observations from Earth-view, little is known about the short-term evolution of CHs and effects on their in-situ signatures.

\begin{figure}
\centering \includegraphics[height=0.7\linewidth,angle=0]{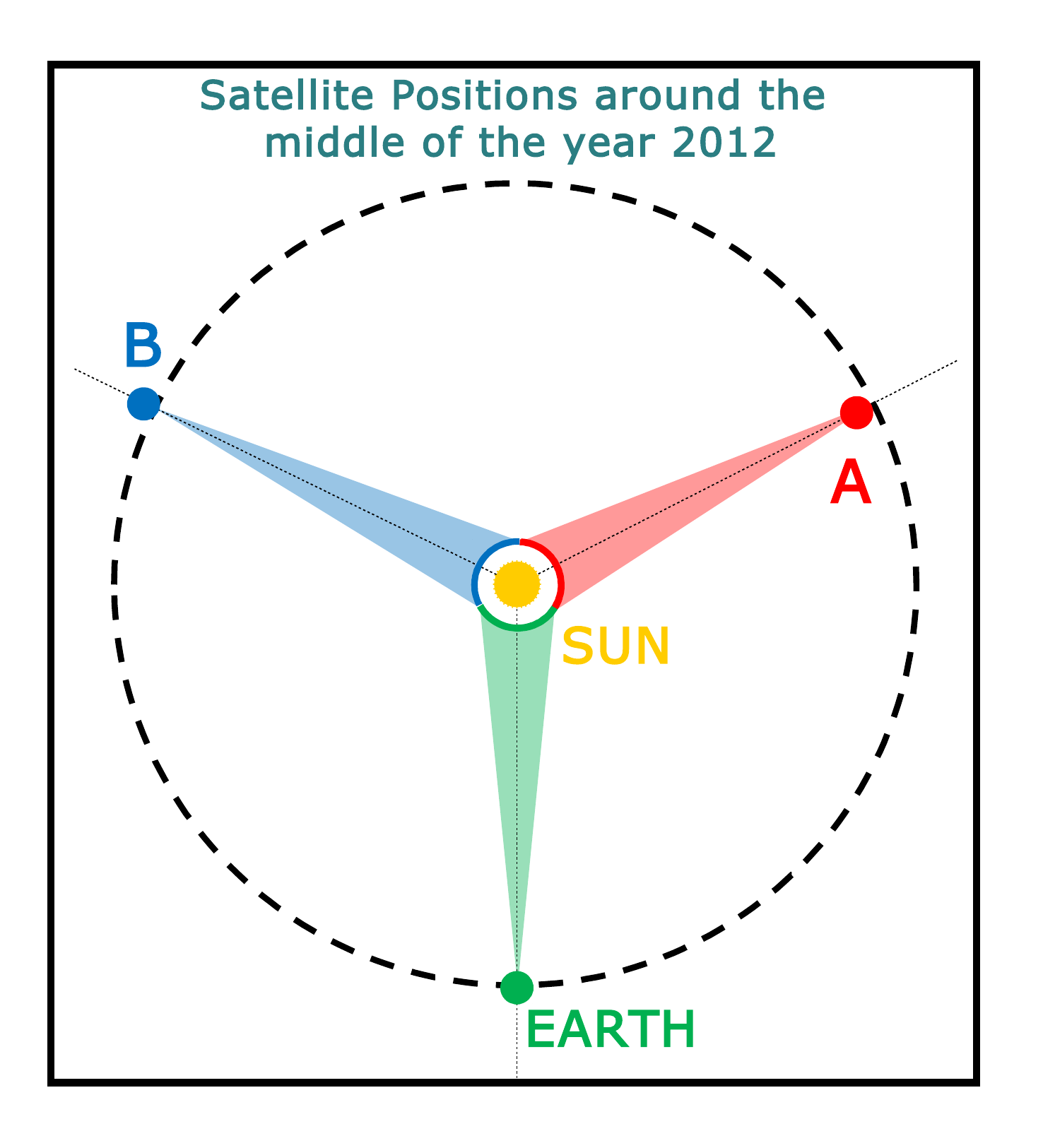}
\caption{Schematic position of STEREO-A, STEREO-B and Earth around the middle of the year 2012 and their estimated field of view.}\label{fig:sir}
 \end{figure}

To increase the temporal resolution and to get more details about the short-term evolution of a CH we investigate combined data from the Solar Dynamics Observatory, SDO \citep{2012pesnell_SDO} and the Solar TErrestrial RElations Observatories, STEREO \citep{2008kaiser_STEREO}. Around the year 2012, the two spacecraft (STEREO-A, and STEREO-B) were separated with each other and with SDO by $120 \arcdeg$ (Fig.~\ref{fig:sir}). This enables us to seamlessly track a CH over $360 \arcdeg$ and to study its entire evolution from its "cradle to grave". We first investigate global CH characteristics like the CH area, intensity, orientation and rotational patterns. The remote sensing observations are complemented by in-situ measurements at the three viewpoints from STEREO-A/-B, the Global Geospace Science Wind satellite \citep{1995acuna_GSS} and the Advanced Composition Explorer, ACE \citep{1998stone_ACE} in the distance of about 1~AU. From that we study the CH associated solar wind plasma and magnetic field parameters.

This study is separated into two papers with the first paper to investigate the CH evolution using combined EUV image data and in-situ measurements from three viewpoints over $360\arcdeg$ of the heliosphere (part~I). The second paper (part~II) covers the magnetic evolution of the same CH on global scales and its fine structure in terms of flux tubes. We note that part~II of the study is restricted to Earth-view (SDO/HMI) only.

\section{Methods} \label{sec:methods}

\subsection{Data} \label{subsec:data}

CHs can be well observed in the emission of highly ionized elements, especially iron (e.g. Fe \textsc{xii}: 193/195\AA~or Fe \textsc{xiv}: 211\AA). Based on the high contrast between CHs and the surrounding quiet corona, the EUV wavelength of 193\AA~from the Atmospheric Imaging Assembly (AIA/SDO) and 195\AA~from the EUV Imager (EUVI/STEREO) is best suited to extract the coronal hole boundaries from the different spacecraft. The observed light is emission from Fe \textsc{XII} ions with a peak response temperature of $1.4-1.6 \cdot 10^{6}$ K \citep{2011aschwanden,2012lemen_AIA}. The data was acquired at a 1 hour cadence through the Joint Science Operations Center (JSOC) and the Virtual Solar Observatory (VSO).

For the in-situ measurements, level 2 data for the three different viewpoints are used. For Earth position, these are the high resolution (5-min) plasma and magnetic field measurements provided by OMNI \url{http://omni- web.gsfc.nasa.gov} and averaged over 1 hour. OMNI data is taken by different satellites (e.g., WIND and ACE) at various positions (e.g., L1) and then propagated to the Earth's Bow Shock Nose.  For the two STEREO positions, we use 1-hour resolution solar wind magnetic field data from IMPACT \citep{2008acuna_IMPACT,2008luhmann_IMPACT} and plasma parameters from PLASTIC \citep{2008galvin_PLASTIC}. 




\subsection{EUV Image Data Reduction and Multi-Instrument Intercalibration} \label{subsec:prep}
To prepare the EUV images for multi-instrument calibration, the images were prepared to level 1.5 and 1 for SDO and STEREO, respectively using standard SSW-IDL routines. For a common size of the images and to enhance the processing speed, full resolution images from SDO (4096$\times$4096) and STEREO (2048$\times$2048) were both resized to 1024$\times$1024 pixels. In order to combine EUV image data from the different instruments, STEREO (195\AA) and SDO (193\AA) are intercalibrated following the procedure described by \citet{2016caplan_LBC_IIT}. To be able to consistently extract CH areas based on their intensity over the solar disk, an inter-instrument intensity normalization is applied and center-to-limb variations (e.g.\,limb-brightening in EUV) are corrected. A robust Limb-Brightening-Correction (LBC) is derived by using unbiased long-term averages of intensity bins. To correct for the different satellite orbits, instruments and instrument conditions, an Inter-Instrument-Transformation (IIT) was applied. The source code for the data based transformations (LBC, IIT) was supplied by the Predictive Science Team in MATLAB and was converted into IDL. For further information about the LBC and IIT, we refer to \citet{2016caplan_LBC_IIT}.




\subsection{CH Extraction and Tracking} \label{subsec:ex}
The CH under study was chosen for reasons that were first of all based on the position of the different spacecraft. For a continuous tracking, STEREO and Earth-view satellites needed to be equally separated, which is only given around the year 2012. During that time, this particular CH was outstanding as it had a long lifespan, a reasonable size and was located at a low-latitudinal position. 




For the extraction of individual CHs on the solar disk, an intensity method with a threshold based on the median intensity of the solar disk is used. As threshold, a value of $35\%$ of the median intensity of the solar disk was applied. Compared to the mean intensity, the median intensity is more robust and stable over time against the influence of bright and dark structures. By applying the threshold method onto EUV images, a binary map is created which is then smoothed by a median filter of 15 pixels (which equals $\sim30$ arcsec in the center of the solar disk) in order to remove fragments, very small dark structures and to smooth the edges of the coronal hole. The binary map is then segmented into the different detected CHs. This threshold based method has already been used and intensively tested by \citet{2014reiss} and \citet{2017hofmeister} for SDO images and by \citet{2012rotter} for EIT images. To reduce projection effects when CHs are close to the limb of the solar disk \citep{1975timothy}, CHs where the center of mass (CoM; see Section~\ref{subsec:global-prop}) is outside of $-45\arcdeg$ and $+45\arcdeg$ degrees longitude measured from the central meridian of the current spacecraft are disregarded. 

The tracking of the CH under study in each image was done using a semi-automatic tracking algorithm. The CH was identified in the first image of the time series and then automatically compared to the following images. 
To identify the CH under study in each image, the CoM and area of all extracted CHs are compared to the forward rotated position of the CH in the previous image. If none or multiple similar CHs where detected, manual input was requested to identify the correct structure. This process was applied to all ($>4000$) images. Manual input was only needed in $<5\%$ of the images.

\subsection{Analysis of Coronal Hole Properties} \label{subsec:global-prop}
Different CH properties, as the de-projected area, the mean intensity, the CoM, and orientation, were calculated from the extracted binary structure. As pixels closer to the solar limb represent a larger area, the binary maps were pixel-wise corrected for the projected area on the spherical solar surface:
\begin{equation}
A_{\mathrm{i,corr}}=\dfrac{A_{\mathrm{i}}}{\cos(\alpha_{\mathrm{i}})},\label{eq:spherical-correction}
\end{equation}
with $A_{\mathrm{i}}$ being the area per pixel and $\alpha_{\mathrm{i}}$ the angular distance to the center of the solar disk. The total area was calculated by summing the corrected area of all CH pixel by
\begin{equation}
A=\sum_{\mathrm{i}}^{\mathrm{N}} A_{\mathrm{i,corr}}.
\end{equation}
The area is given in square kilometers, $\mathrm{km^{2}}$. Using the de-projected area of the CH, the CoM is calculated pixel wise by weighing the pixels based on their area. We further calculate the longest diagonal of the CH that passes through the CoM, from which we define the orientation ($\phi$) of the CH as the counterclockwise angle of its longest diagonal with respect to the solar equator. The mean intensity of the CH is calculated by 
\begin{equation}
\bar{I}=\dfrac{1}{N}\sum_{\mathrm{i}}^{\mathrm{N}} I_{\mathrm{i}}.
\end{equation}
With $I_{\mathrm{i}}$ being the intensity value of each CH pixel. 


\subsection{Solar Wind Plasma and Magnetic Field Parameters} \label{subsec:sw-data}
To identify from the in-situ solar wind data high-speed streams associated with the CH under study, we have to take into account the propagation time from Sun to Earth. A CH that is centrally located on the solar disk shows a strong increase in density and magnetic field after about 1--2 days, and its peak in the solar wind plasma speed on average after 2--6~days \citep[e.g.,][]{2007vrsnak}. We therefore use the central meridian passage (CMP) of the CoM as reference and open a time-window in the in-situ data set of $-4$ to $+8$ days (minus to account for E-W elongated CHs). After manually identifying the stream, the values of the solar wind plasma (peak velocity and total perpendicular pressure) and magnetic field (peak in the total magnetic field at compression region, and magnetic field at the time when the speed peaks) were manually extracted.  The total perpendicular pressure (sum of magnetic pressure and perpendicular plasma thermal pressure) was calculated using the formula given by \citet{jian06}: P$_{\mathrm{t}}=B^2/(2\mu_{0}) +\sum_{j} n_{j}kT_{\mathrm{perp},j}$, where $j$ represents protons, electrons and $\alpha$ particles. For both, OMNI and STEREO data, we assume a constant electron temperature of $130000$ Kelvin and a constant ratio of $\alpha$ particles to protons of $4\%$. To account for the temporal extension of the stream, we averaged the in-situ data of each parameter over $\pm 12$ hours around the manually extracted peak value.

\begin{figure*}
 \plotone{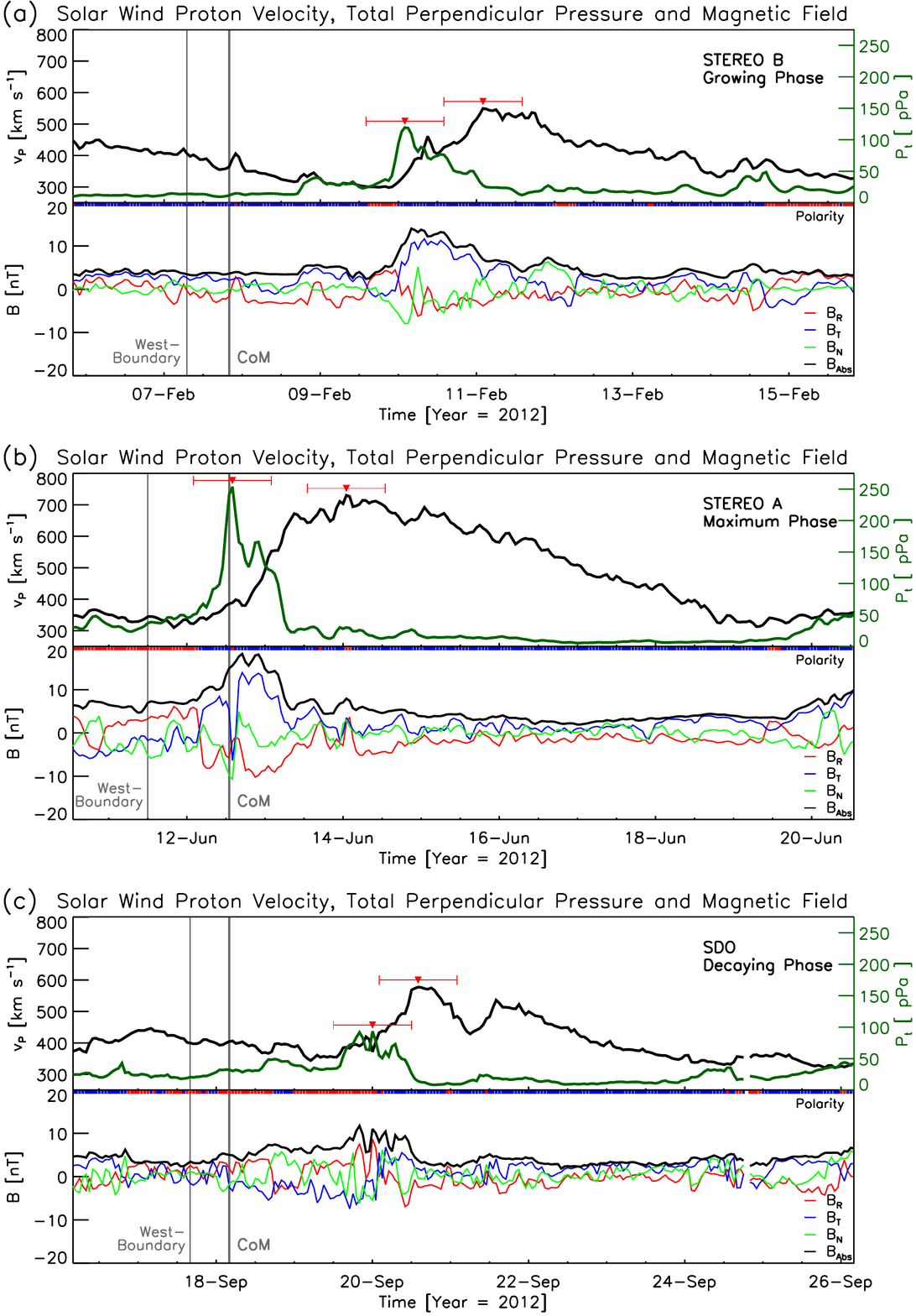}
 \caption{Examples of in-situ signatures of the solar wind as observed by STEREO-B,  STEREO-A and OMNI as the CH under study was observed near the center of the solar disk by STEREO-B, STEREO-A and SDO, respectively. The panels a--c represent different stages in the CH evolution. The top panels show the solar wind bulk velocity (black) and total perpendicular pressure (green) and the bottom panels show the magnetic field components (RTN and absolute value). The red-blue bar below the velocity-pressure plot indicates the magnetic field polarity of the solar wind, with red representing positive and blue negative (=CH) polarity. The gray vertical lines are the time when the CH makes an CMP with the CoM and the west boundary. The red bars indicate the manually extracted peak (triangle) and the $\pm 12$ hours interval over which the parameter was averaged.}\label{fig:insitu}
 \end{figure*}

Figure~\ref{fig:insitu} shows three examples of the in-situ data that was used. Within the defined time-window, after the marks for the CMP of the west boundary and the CoM (gray vertical lines), we looked for typical HSS signatures using the criteria as given by \citet{jian09}. A HSS signature can usually be identified by a pile up of the total perpendicular pressure  and a steep peak (shock) in the proton density followed by a rise in the bulk velocity that usually peaks a couple of hours/days later and then slowly decreases (upper panel). The compression affects the density, pressure as well as the magnetic field which usually also is sheared (lower panel). Also the polarity of the magnetic field of the HSS has to be considered as it has to match the polarity of the source region, i.e.,\,the CH, which is found to be negative for our event under study \citep[see~][]{2018heinemann_paperII}. The HSS polarity was calculated using the formula given by Equation 1 in \citet{2002neugebauer} and is marked as red and blue bar below the velocity-pressure plot. The red bars in each plot mark the manually extracted peak (triangle) and the $\pm 12$ hours interval over which the parameter was averaged. For each disk passage of the CH observed in the three different satellites (i.e.,\,29 disk passages in total) we extract one HSS signature with its respective properties. 

Possible CME signatures in the HSS were cross-checked by using ready-catalogs maintained by Richardson \& Cane\footnote{\url{http://www.srl.caltech.edu/ACE/ASC/DATA/level3/icmetable2.htm}} for ACE (see \citealt{2010richardson_RC-list} for a description of the catalog) and for STEREO from L. Jian\footnote{\url{ftp://stereodata.nascom.nasa.gov/pub/ins_data/impact/level3/STEREO_Level3_ICME.pdf}} (see \citealt{jian18}, for a description of the catalog). From a total of 29 data points we excluded five from the analysis: March 5, 2012 because it was a CME-HSS interaction event and the HSS signatures could not be properly separated from the CME; August 5, 2012 because only a CME signature is visible; August 14, 2012, August 22, 2012 and October 14, 2012 were excluded as no clear HSS signature could be identified associated with the CH under study.




To compare and relate the CH evolution, observed in the remote sensing data, to the in-situ data, the area of the CH was processed. Based on the time of the central meridian passage (CMP) of the CoM, the area was averaged over $\pm 18$ hours ($A_{CH,n}$) to derive a CH area to associate with the in-situ measured HSS signature:

\begin{equation}
A_{\mathrm{SW}}=\dfrac{1}{N}\sum_{\mathrm{n}} A_{\mathrm{CH,n}}.
\end{equation}

\startlongtable
\begin{deluxetable}{ll|l}
\tablecaption{Overview of Parameters Defined in Section~\ref{sec:methods} \label{tab:methods}}
\tablehead{
\colhead{Parameter} & \colhead{Definition} & \colhead{Description}
}
\startdata
$A_{\mathrm{CH}}$ & $=\sum_{\mathrm{i}} A_{\mathrm{i,corr}}$ & CH Area \\
$I_{\mathrm{CH}}$ & $=\tfrac{1}{N}\sum_{\mathrm{i}} I_{\mathrm{i}}$ & CH Mean EUV Intensity \\
$\phi _{\mathrm{CH}}$ &  & CH Orientation \\
$v_{\mathrm{SW}}$ &  & SW Peak Bulk Velocity \\
$\rho_{\mathrm{SW}}$ &  & SW Number Density \\
$B_{\mathrm{SW}}$ &  & SW Magnetic Field Strength \\
$P_{\mathrm{t}}$ & $=B^2/(2\mu_{0})$ & SW Total Perpendicular \\
& $  +\sum_{j} n_{j}kT_{\mathrm{perp},j}$ & Pressure  \\
$A_{\mathrm{SW}}$ & $=\tfrac{1}{N}\sum_{\mathrm{n}} A_{\mathrm{CH,n}}$ & Processed CH Area \\
\enddata
\end{deluxetable}

\subsection{Visualization and Correlation} \label{subsec:vis}
\begin{figure*}
\centering \includegraphics[height=1\linewidth,angle=90]{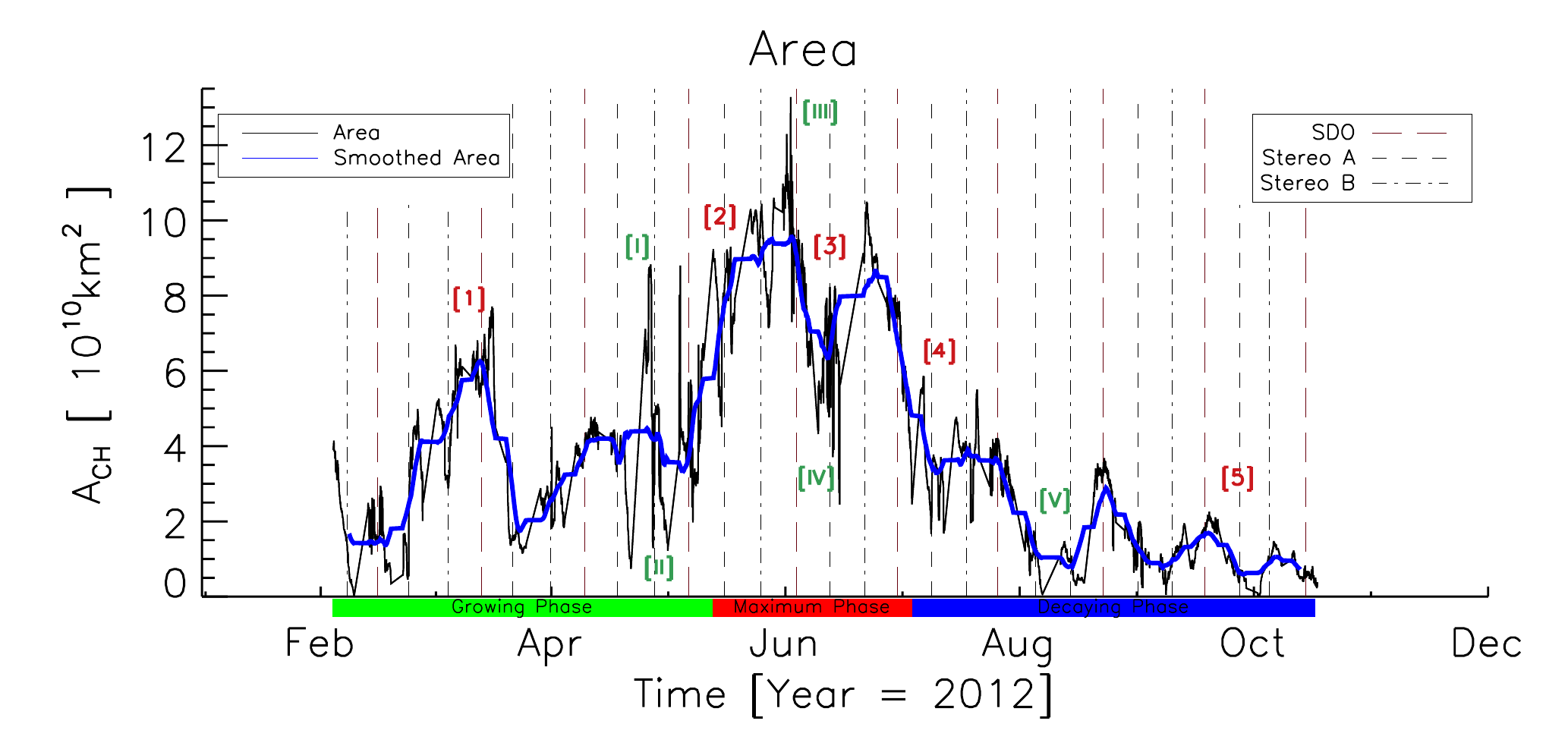}
\caption{Evolution of the CH area (black) with smoothed curve (blue). The red numbers mark evolutionary features described in the text and the green roman numbers mark outliers. The vertical lines represent the observing spacecraft as described in Subsection~\ref{subsec:vis}.}\label{fig:area}
\end{figure*}

To visualize the different spacecraft, the CH evolution plots show three different vertical lines. The \textit{dashed} line marks the STEREO-A spacecraft, the \textit{dashed-dotted} line STEREO-B, and the \textit{dark red dashed} line SDO. Each of those lines represents the CMP, hence, the time when the CoM of the CH passes the central meridian of the respective spacecraft. These are featured in the Figures~\ref{fig:area},~\ref{fig:int}(a),~\ref{fig:mov} and~\ref{fig:insitu4}(a),~(c),~(d).

For correlating various parameters, the Pearson correlation coefficient and the Spearman correlation coefficient including the confidence intervals (CIs) have been calculated by bootstrapping \citep{efron93_bootstrap,efron1979_bootstrap} the dataset with over $10^{5}$ repetitions. For each repetition, a subset with replacement was drawn from the initial set, with each data point in this subset coming from a Gaussian distribution of itself plus its standard deviation. From the resulting subset the correlation coefficients were calculated. The given correlation coefficients are the mean values of all repetitions. The confidence intervals (CI) for the correlation coefficients ($90\%,~95\%,~99\%$) were calculated using the respective quantiles (e.g.\,for the $95\%$ CI the quantiles are $2.5$ and $97.5 \%$). A summary of all results and statistical parameters is given in Table~\ref{tab:corr} in the Appendix.    

\section{Results on Coronal Hole Properties} \label{sec:CH-prop}
\begin{figure*}
 \plotone{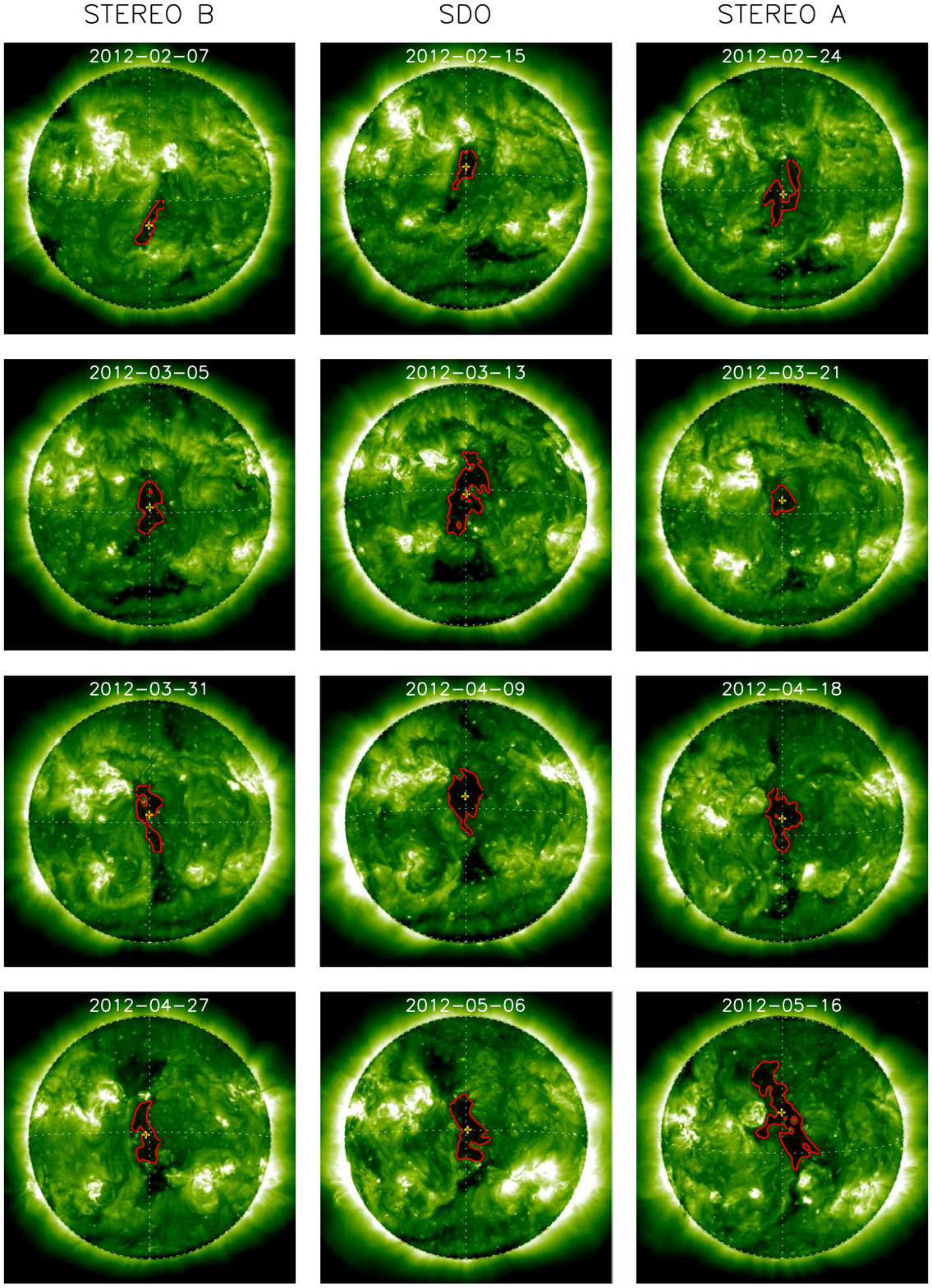}
 \caption{Evolution of the CH during the time frame from February 07, 2012 to May 16, 2012. The red contours represent the boundaries of the CH. The yellow$-$black cross is the location of the CoM. The left panels show images taken by STEREO-B, the central panels are images by SDO and the right panels images by STEREO-A. An animation of Figures ~\ref{fig:evo-plot1} -~\ref{fig:evo-plot3} is online available.}\label{fig:evo-plot1}
 \end{figure*}

\begin{figure*}
 \plotone{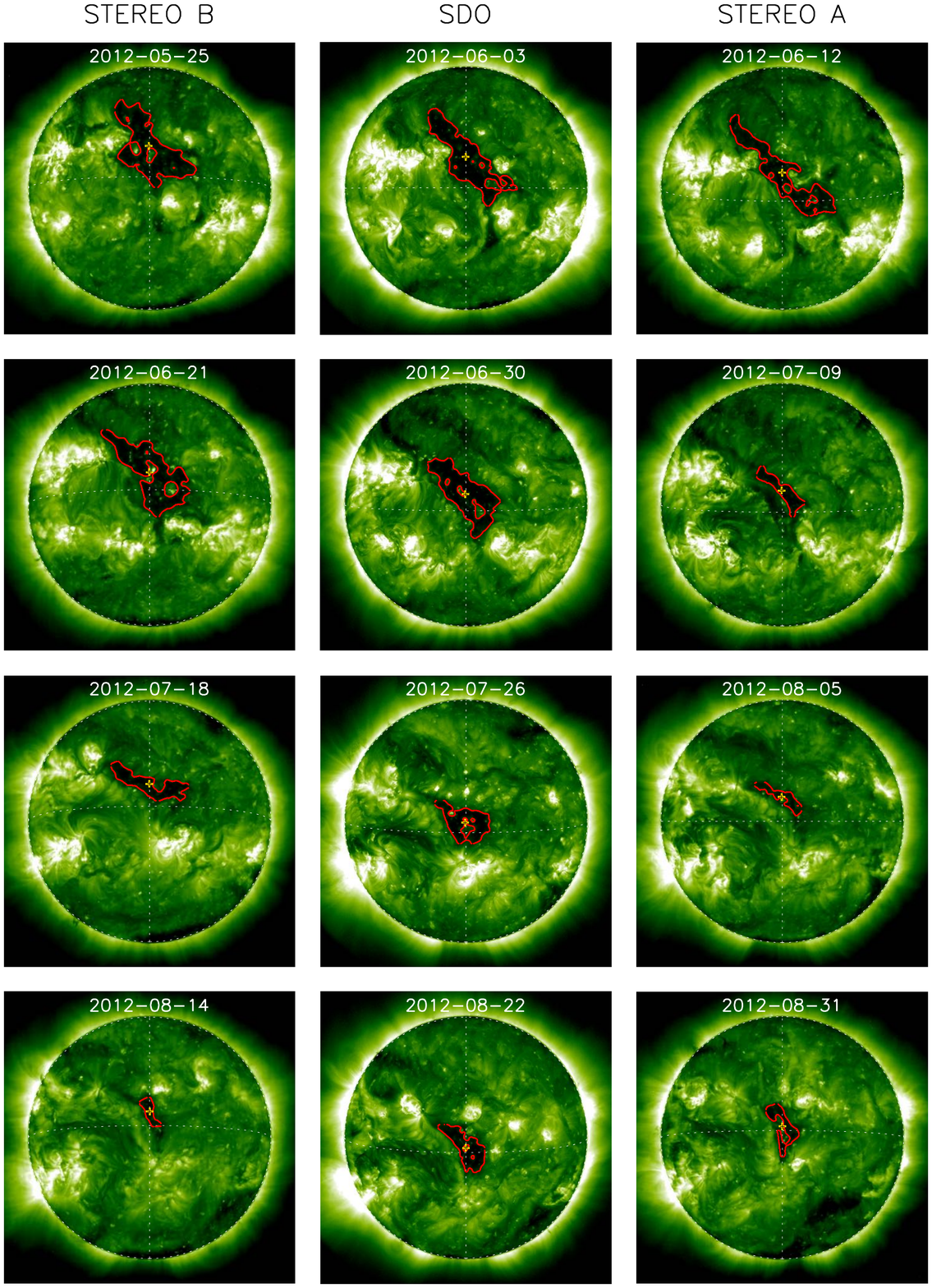}
 \caption{Same as Figure~\ref{fig:evo-plot1} but from May 25, 2012 to August 31, 2012.}\label{fig:evo-plot2}
 \end{figure*}

\begin{figure*}
 \plotone{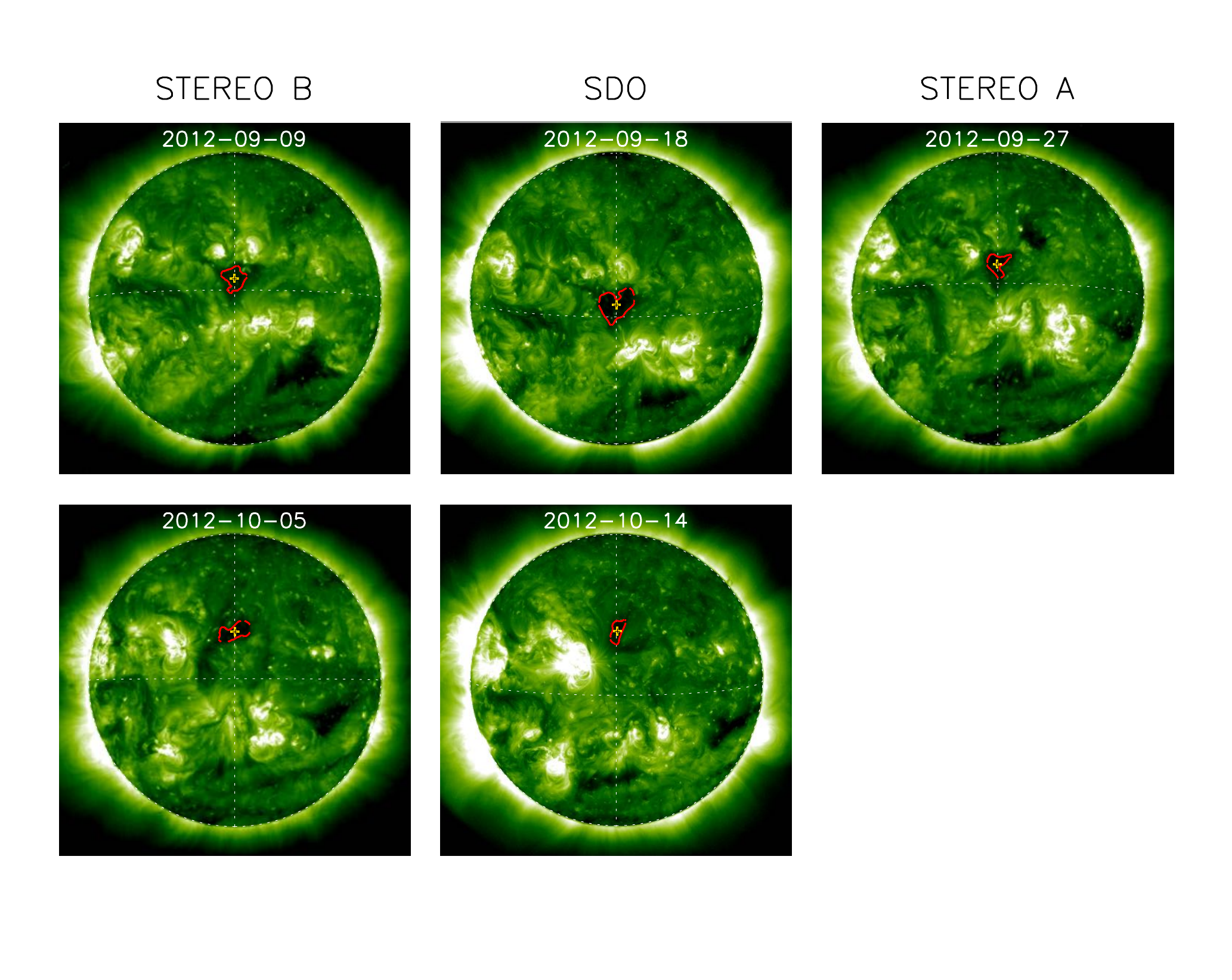}
 \caption{Same as Figure~\ref{fig:evo-plot1} but from September 09, 2012 to October 14, 2012.}\label{fig:evo-plot3}
 \end{figure*}

Figures ~\ref{fig:evo-plot1} -~\ref{fig:evo-plot3} show an overview of the evolution (snapshots with a cadence of 9 days) of the CH at the CMP, observed in SDO and STEREO EUV image data over its entire lifespan from February 4, 2012 to October 17, 2012. An animation of the evolution is available online.


\subsection{Coronal Hole Area and Intensity} \label{subsec:area}
 
\begin{figure}
 \centering \includegraphics[width=1.05\linewidth,angle=0]{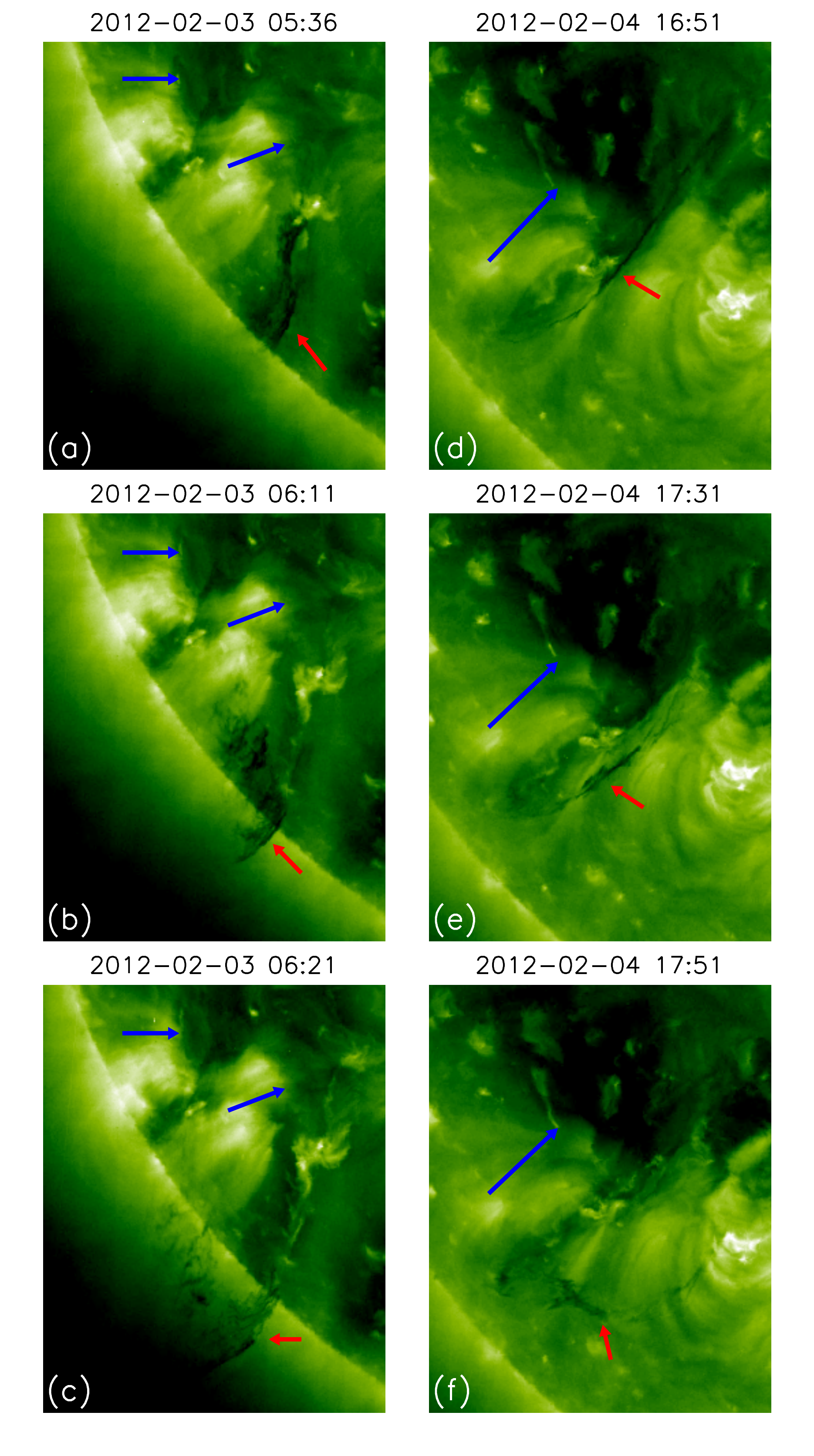}
 \caption{The formation of the CH, initiated by two filament eruptions. The red arrows point to the filament and the blue ones towards the CH that forms. Panel a--c show three snapshots in the eruption of the first filament on February 03, 2012 and panels d--f show three snapshots of the second filament eruption.}\label{fig:birth}
 \end{figure}

The CH under study became continuously detectable starting with February 3--4, 2012 after two consecutive filament eruptions that have evacuated material and opened the magnetic field lines (see Fig.~\ref{fig:birth}). This is observed in STEREO-A off-limb and in STEREO-B very close to the limb. Signs for the emerging of a structure of reduced intensity can already be seen before in STEREO-A and SDO, however no continuous extraction of the structure was possible nor could we associate a HSS in the in-situ data to this structure. Thus we consider the double filament eruption event as the "formation" of the CH, and as such it marks the start time of our analysis. Because the position of the formation of the CH is close to the solar limb, a detailed analysis of the opening process was not feasible.

Figure~\ref{fig:area} shows the evolution of the CH area as measured from the three different spacecraft. The black line represents the calculated area from the extracted CHs, the blue line is a smoothed curve (running median of 100 images $\approx~3$ days) to make the trend visible and to remove outliers. Extreme outliers are marked with the green roman numbers and are caused by the extraction algorithm that gives a false merging and splitting of the CH with fractions of other low intensity areas close to the CH. From the derived profile, we find that the area evolves with a pattern that can be divided into three distinct phases. 

The \textit{growing phase} is related to an increase in area ($>6\cdot10^{10}~$km$^{2}$) from February until the middle of March ([1]) where the area decreases to $\sim 2\cdot10^{10}~$km$^{2}$. From observations, we obtain that this decrease may be explained by interchange reconnection with a small emerging active region near the southern part of the CH and/or by overlying coronal loops and stray light from active regions nearby which obscure the CH. After this drop, again a steady increase is observed until the area reaches the \textit{maximum phase} ($> 9\cdot10^{10}~$km$^{2}$) around May 13, 2012 ([2]). This maximum is very stable (with only slight variations, e.g.\,the dip around the start of June ([3]) and lasts until around July 03, 2012 ([4])). After the maximum\footnote{The transition dates for the maximum phase of the CH are based on a threshold value for the CH area of $\sim 6\cdot 10^{10}$ km$^{2}$.}, we observe the \textit{decaying phase} during which the CH area steadily declines to $< 2\cdot10^{10}~$km$^{2}$ ([5]) until the small CH cannot be detected anymore after October 17, 2012.

\begin{figure}
\centering \includegraphics[height=1\linewidth,angle=90]{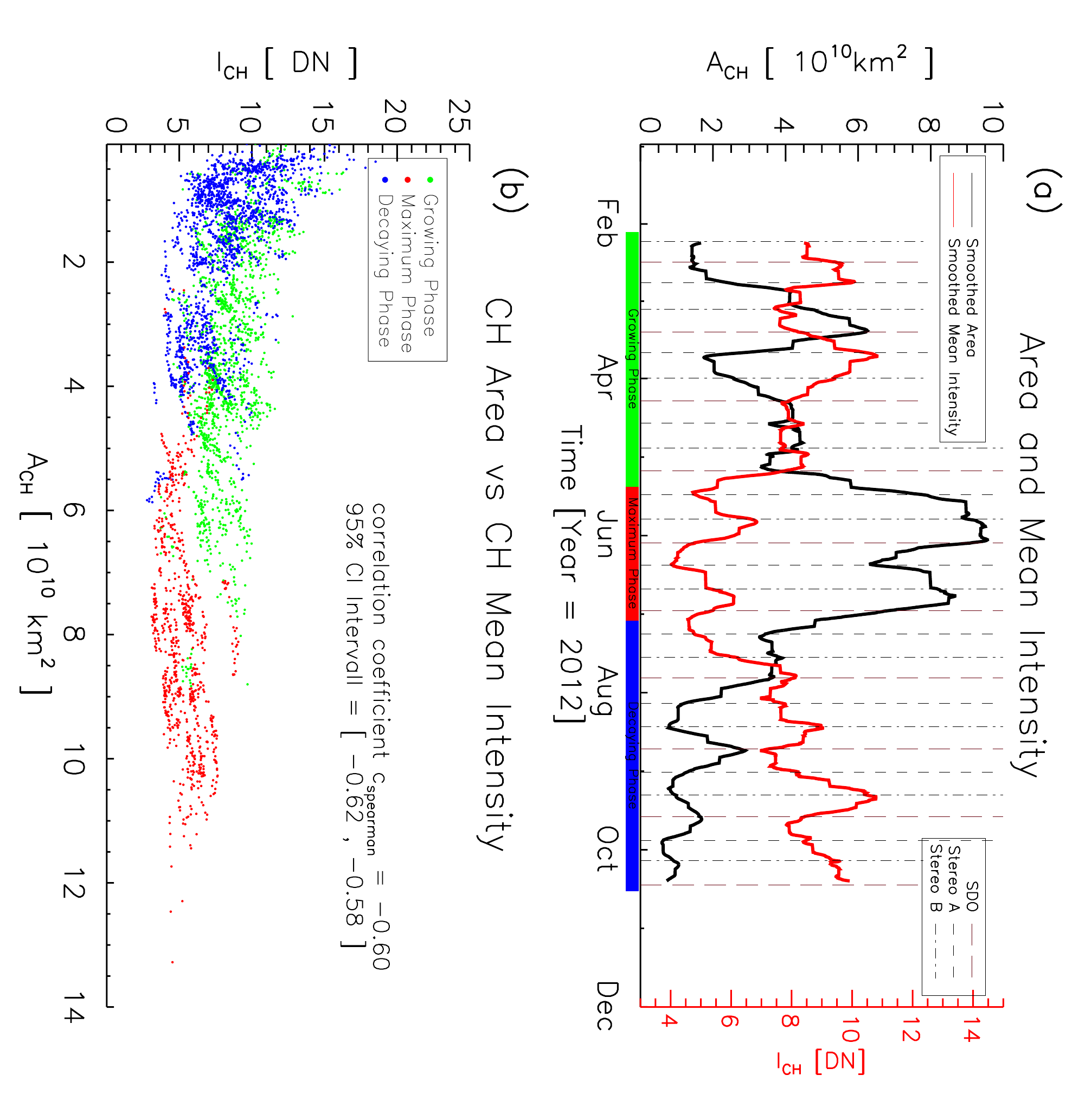}
\caption{Panel (a) shows the evolution of the CH mean intensity (smoothed, red) and the area (smoothed, black). The vertical lines represent the observing spacecraft as described in Subsection~\ref{subsec:vis}. In Panel (b) the CH area is plotted against its mean intensity. The 3 different colors represent the evolutionary phases: green: the growing phase, red: the maximum phase and blue: the decaying phase.}\label{fig:int}
 \end{figure}

In the following, the 3-phase evolution found in the area is used as a reference, and we investigate the other CH parameters in comparison to that result. Figure~\ref{fig:int}(a) shows the smoothed mean intensity together with the smoothed area. We find that the smoothed mean intensity of the CH changes by about a factor of $2.5$ and that it is anti-correlated with the area evolution. The minimum of intensity corresponds with the maximum in the area. In Figure~\ref{fig:int}(b) the CH areas are plotted against their respective mean intensities. The data points of the different evolution phases are color-marked (green: growing phase, red: maximum phase, blue: decaying phase). We find a good (anti-)correlation, expressed through a Spearman correlation coefficient of $-0.60$ with a $95\%$ confidence interval of $[-0.58,-0.62]$. The different stages in evolution occupy different areas in the plot, separating the phases most notably for the growing and maximum phase. The decaying phase is more scattered showing some overlap with the growing phase.

\subsection{Coronal Hole Motion Patterns} \label{subsec:mov}

\begin{figure}
\centering \includegraphics[height=1\linewidth,angle=90]{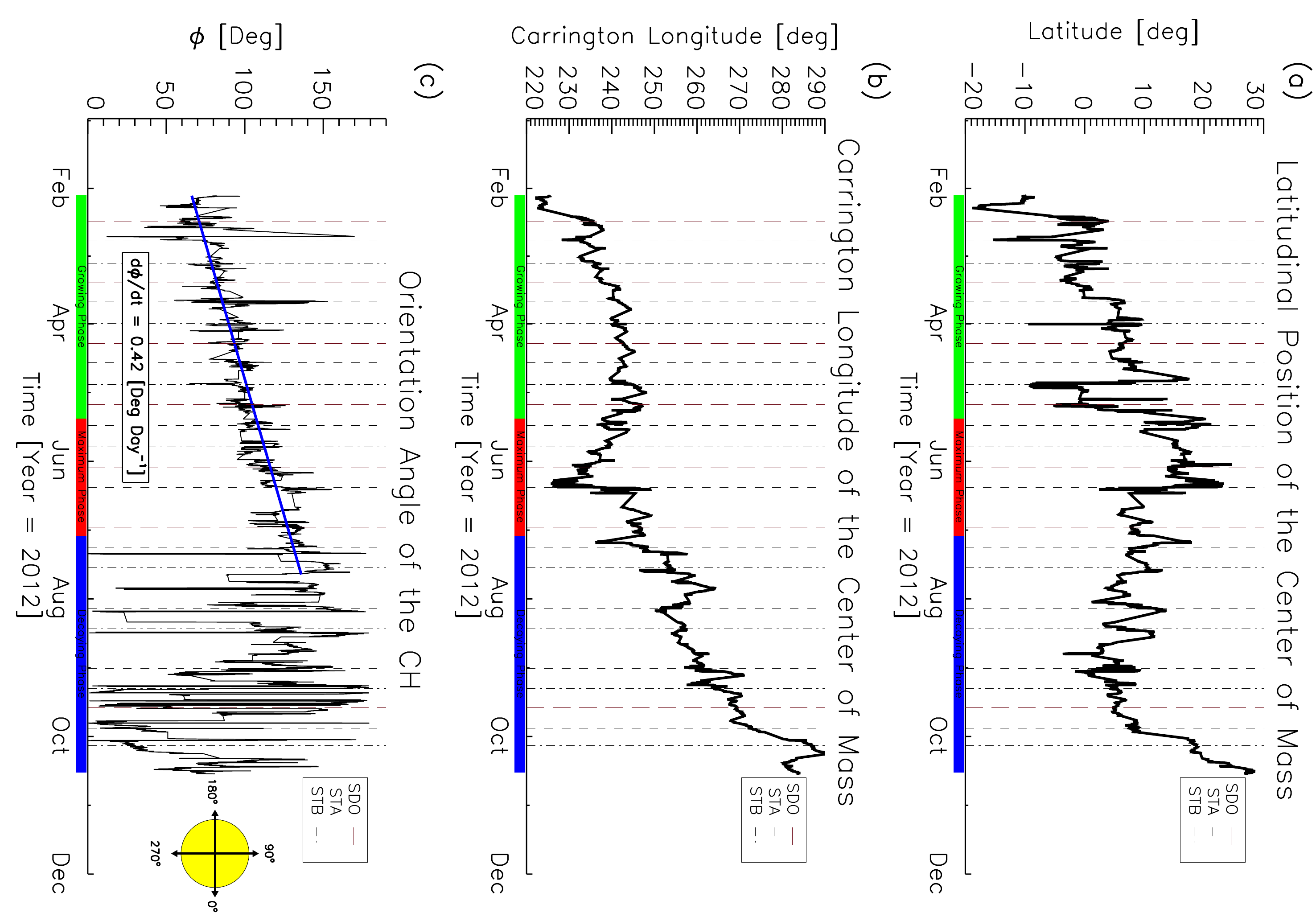}
\caption{Motional behavior of the CH: (a) latitudinal movement of the CH; (b) longitudinal motion of the CH; (c) orientation angle of the CH together with a linear fit (blue) for the time where the angle can be derived unambiguously. The vertical lines in a--c represent the observing spacecraft as described in Subsection~\ref{subsec:vis}. }\label{fig:mov} 
 \end{figure}

Figures~\ref{fig:mov} shows different properties of the CH motion patterns. The latitudinal position over time (Fig.~\ref{fig:mov},a), the velocity of the CoM in form of the Carrington longitude (Fig.~\ref{fig:mov},b) and the CH orientation (Fig.~\ref{fig:mov},c).  
 
The latitude of the CoM is located first in the southern hemisphere at around $20\arcdeg$ and moves successively upwards into the northern hemisphere until the maximum phase. A subsequent slow movement towards the equator coincides with the decay of the area. With September a fast northward movement ($>25\arcdeg$) is derived, before the CH cannot be observed anymore. Note, that the motion of the CoM carries no information of the shape and extension of the CH,.

In Figure~\ref{fig:mov}(b) the angular rotation speed of the CoM of the CH with respect to the Carrington rotation is shown. A constant Carrington longitude would refer to a rotation speed matching the Carrington rotation $\omega_{\mathrm{Carr}}=14.18~\arcdeg/$day \citep{2012ridpath}. We obtain that the rotation speed changes over the lifetime of the CH (between $13.7~\arcdeg/$day and $14.8~\arcdeg/$day). We observe a decrease in angular velocity until the middle of the maximum phase. There, a sudden increase in the angular velocity can be observed, which is most likely a result of the recession of the area in the northern part of the CH that causes the CoM to shift towards the equator. Thereafter a constant angular rotation speed can be observed.

Figure~\ref{fig:mov}(c) shows the change in the orientation of the CH. The orientation is defined as the counterclockwise angle of the longest CH diagonal to the equator. We find a good correspondence to the three phases as observed in the CH area, especially visible is the transition from the maximum to the decaying phase. A linear fit from February, 2012 until July, 2012 reveals a near constant counterclockwise rotation around its CoM with $0.42~\arcdeg/$day after a starting angle of $\sim70\arcdeg$. This rotation can be observed until the end of the maximum phase, after that, coinciding with the decrease in CH area, the shape of the CH starts to become roundish. For a roundish CH, the orientation is not a well defined parameter and therefore gives no reasonable results but causes strong fluctuations in the orientation angle derived.

\section{Solar Wind Properties} \label{sec:SW-prop}

Many studies already have shown that a linear relation between the CH area and the in-situ measured solar wind bulk speed exists (e.g., \citealt{1973krieger, 1976nolte,2007vrsnak, 2017tokumaru,2018hofmeister}). With STEREO we are able to investigate in more detail how the different stages of a CH evolution affect the solar wind properties measured at 1AU.  

\begin{figure*}
\centering \includegraphics[height=1\linewidth,angle=90]{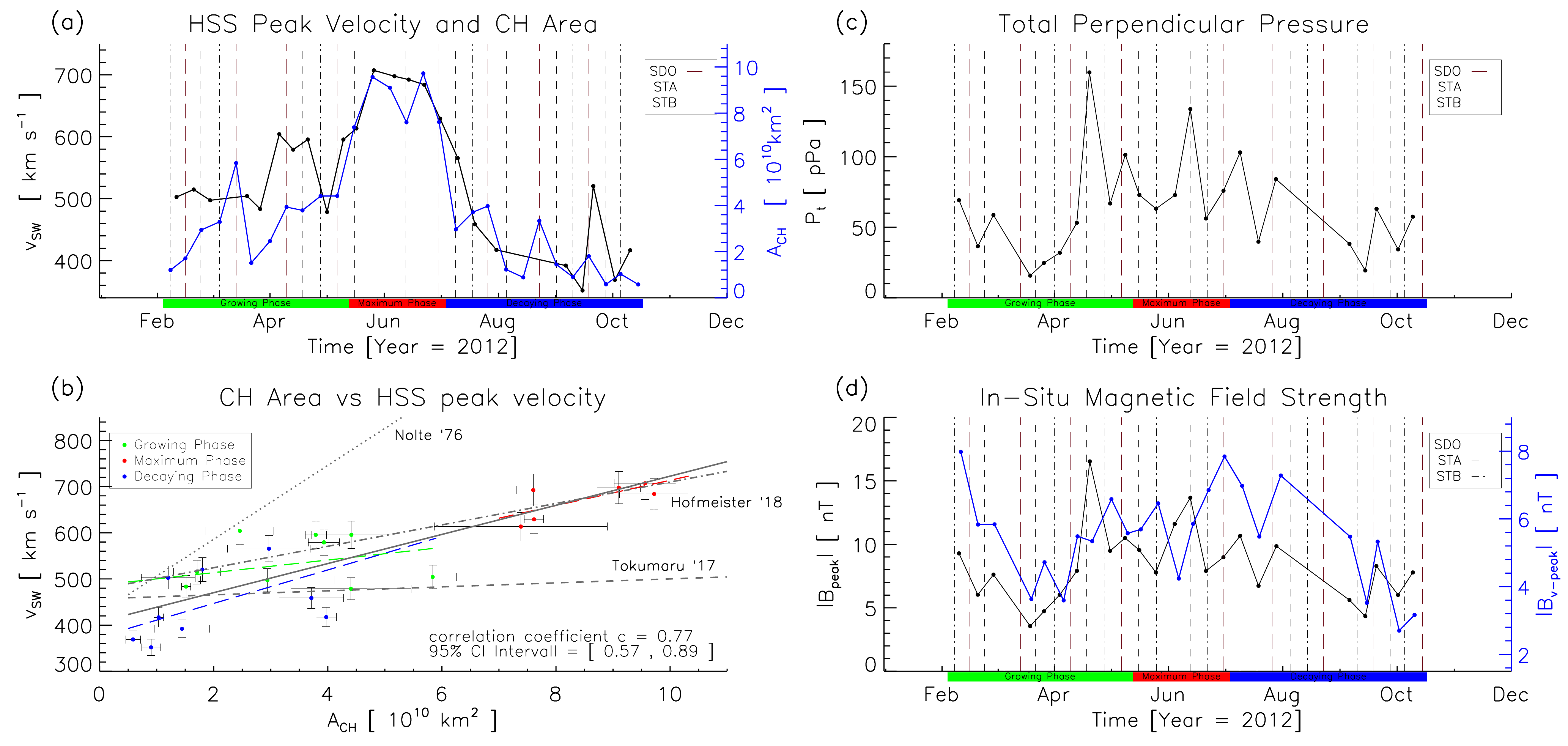}
\caption{(a) Time evolution of the solar wind peak velocity (black) of the HSS associated with the CH together with the evolution of the CH area (blue). (b) HSS peak velocity plotted against the CH area. The three different colors represent the three different evolutionary phases (see legend). The colored lines represent a linear fit for each respective phase. The gray lines (dotted, dashed and dash-dotted) show relations found by the studies of \citet{1976nolte,2017tokumaru,2018hofmeister}. The solid gray line represents the linear relation found in this study. (c) Total perpendicular pressure of the compression region. (d) Evolution of the peak magnetic field strength of the compression region and the magnetic field at the velocity peak. The vertical lines represent the observing spacecraft as described in Subsection~\ref{subsec:vis}}\label{fig:insitu4}. 
 \end{figure*}

In Figure~\ref{fig:insitu4}(a) the solar wind peak velocity and CH area evolution is shown. The peak velocity varies from $350~$km/s during times when the CH area is small up to $>700~$km/s during the time of the maximum in the area. As shown in Figure~\ref{fig:insitu4}(b) we find a strong correlation between HSS peak velocity and CH area with a Pearson correlation coefficient of $0.77$ with a $95\%$ CI of $[0.57,0.89]$. The most clear relation between CH area and solar wind speed is derived for the maximum phase, whereas it is weaker for the growing and decaying phase. The decaying phase is related to a lower speed compared to the growing phase. Interestingly, CH areas during the decaying phase are related to a wide spread of speeds ($350-550$ km/s) while for the growing phase we observe speeds in the range of $500-650$ km/s and during the maximum phase of $600-720$ km/s. Comparing our results with previous studies \citep{1976nolte, 2017tokumaru,2018hofmeister} we are in good agreement with \citet{2018hofmeister}, but derive significant deviations from the slopes of the other two studies.

In Figure~\ref{fig:insitu4}(c) the evolution of the total perpendicular pressure at the CIR compression region is shown. We find significant differences in the values between the different spacecraft, especially STEREO-A. However, some systematic evolution can be derived but not in accordance with the three phases from the area evolution. At the start and end of the CH lifetime, we find lower values ($20-80$ pPa) than [during and] around the maximum phase from May to August, 2012 ($70-160$ pPa). 

Figure~\ref{fig:insitu4}(d) shows the evolution of the peak magnetic field strength at the CIR compression region and at the time of the velocity peak. Similar as for the pressure (P$_{t}$), we observe an evolutionary trend of lower values for the beginning and end of the lifetime ($3-10$ nT and $2-5$ nT respectively) and higher values from May to August, 2012 ($8-17$ nT and $4-8$ nT respectively). However, this trend is not in accordance with the three phases from the area evolution.

\section{Discussion} \label{sec:dis}
Using multi-viewpoint data, we have investigated the evolution of a long-lived low-latitude coronal hole over its entire lifetime of more than 10 full solar rotations. With combined SDO and STEREO data we could seamlessly track and investigate the evolution of the CH in EUV imagery as well as in in-situ properties of the associated HSS. Our main finding is that the CH evolves in three distinct phases: the growing phase, the maximum phase and the decaying phase.

\subsection{3-Phase Evolution}
Most prominently the 3-phase evolution is derived in the CH area. The \textit{growing phase} is mainly defined by a steady increase in area with a duration of about three months, until the transition to the \textit{maximum phase} takes place. Compared to the other phases, the \textit{maximum phase} is characterized by a larger CH area (Fig.~\ref{fig:area}), lower mean EUV intensity (Fig.~\ref{fig:int}), a steady high solar wind speed of the associated HSS and a stronger density/magnetic field compression of the CIR (Fig.~\ref{fig:insitu4}). The parameter values are rather stable during the \textit{maximum phase} but only of limited duration of around one month (the shortest of the three evolutionary phases). Around July 03, 2012 a significant drop in CH area and in-situ solar wind speed heralds the last phase, the \textit{decaying phase} in the CH lifetime. The decay continues for more than 3 months during which the CH area decreases and finally cannot be unambiguously identified anymore.

In each phase the CH exhibits a different behavior, especially noticeable is the difference between the growing and the decaying phase. Though the CH has in both phases similar values of area and similar solar wind plasma and magnetic field values, the solar wind plasma stream emanating from the decaying CH behaves more erratic compared to the CH growing phase. This leads us to the assumption that the physical processes that are responsible for the opening and for closing of a CH have a different influence on the HSS peak velocity \citep[see also ][]{temmer18}.

\subsection{Motional Behavior}
The CH main latitudinal movement direction is northwards with a slight decrease in CoM latitude at the end of the maximum phase and the start of the decaying phase, which seems to be caused by the asymmetric (north-south direction) decline of the area (Fig.~\ref{fig:mov}). Structural changes of the CH, hence, changes in the CoM and its latitudinal movement, are supposed to be strongly related to the rearrangement of the global magnetic field structure (\citealt{2002bilenko} and \citealt{2016bilenko}).

The CH main axis exhibits a counterclockwise rotation (Fig.~\ref{fig:mov},c) which might be caused by the differential rotation of the Sun acting differently on different parts of the CH. The CH is located mostly in the northern hemisphere stretching from the equator up to latitudes of $30\arcdeg$ and higher. The faster rotation rates at the equator and the lower rotation rates at higher latitudes deform the CH, which becomes apparent as a counterclockwise rotation. This effect is especially noticeable when the CH has an elongated shape in the north-south direction. This finding shows that an extended CH does not rotate rigidly as a static body, but rather different parts have different rotation speeds, faster at the equator and slower at higher latitudes. A coronal differential rotation in CHs has also been found by \citet{1995insley}.

\subsection{In-situ solar wind plasma and magnetic field signatures}
Solar wind HSSs and related CIRs are a major contributor to space weather, and due to their recurrence more predictable than impulsive and sporadic space weather events (e.g.\,CMEs). With the well derived linear relation between CH areas and HSS peak speed \citep[e.g., ][]{1976nolte,2007vrsnak,2012rotter,2017tokumaru,2018hofmeister}, a forecasting seems to be straightforward. In this study, we show that these correlations are well related to the evolution of the CHs (Fig.~\ref{fig:insitu4}). During the CH area growing phase, the peak velocity of the associated HSS steadily increased from $500$ km/s to $> 680$ km/s in the maximum phase. With the decrease in area in the decaying phase, the speed drops to below $400$ km/s. The correlation, as shown by the Pearson correlation coefficient is 0.77 with a $95\%$ CI of $[0.57,0.89]$. 

From our regression analysis we obtain a linear relation with a slope of $31.5\pm2.2$ km~s$^{-1}/(10^{10}$ km$^{2}$). Comparing, \citet{2017tokumaru} derived a smaller slope with $4.31\pm0.15$ km~s$^{-1}/(10^{10}$ km$^{2}$) and \citet{1976nolte} a higher slope with $80\pm2$ km~s$^{-1}/(10^{10}$ km$^{2}$). We are in basic agreement with the study by \citet{2018hofmeister} with a slope of $23.2\pm4.5$ but a higher y-intercept. These strong variations in the slope in different studies may be explained by the different CH data sets used. In a recent statistical study by \citet{2018hofmeister} a further dependence of the measured HSS speed (alternatively, of the measured slopes) on the co-latitudinal position between the spacecraft and the solar CH was found. However, since our case study only covers a low-latitude CH within the co-latitudinal range of $\pm 20 \arcdeg$, such a correlation is not visible. Studies with only near equatorial CHs are expected to yield a different slope than those covering CHs over all latitudinal ranges. This assumption also applies to this study of the evolution of one CH. Nevertheless, as could be shown here, the degree of correlation is likely dependent on the evolutionary phase of the CH. We find during the growing and decaying phase similar slopes but with different base levels (lower speed and larger spread for the decaying phase). During the maximum phase the linear correlation is the same as the overall correlation.

The in-situ measured density peak, magnetic field and therefore total perpendicular pressure of the interaction region and the magnetic field measured at the speed peak show a general evolutionary trend, with higher values during the maximum of the CH lifetime than during its beginning and end (Fig.~\ref{fig:insitu4}). For the total perpendicular pressure, we find large variations between the three spacecraft which may be caused by a) instrumental effects (sensitivity, age, type) or b) the 3-dimensional structure of the HSS. However, we find no direct dependence on the co-latitude of the spacecraft to the CoM of the CH. We suppose that the plasma and magnetic field compression between fast and slow solar wind is likely influenced by the interplay between the CH structure itself (shape of the CH and its changes) and the ambient structures (nearby active regions and their activity, but also global changes in the magnetic field).

\section{Summary \& Conclusions} \label{sec:sum}
In the year 2012, the position of the two STEREO spacecraft combined with SDO, gave us the unique possibility to observe and study the long term evolution of a CH in full $360\arcdeg$ view and the associated HSS. Our major findings are the following:\\

\renewcommand{\labelenumi}{\theenumi.}
\begin{enumerate}
\item We find that the CH shows a 3-phase evolution: the growing phase, the maximum phase and the decaying phase. The three phases last for three, one, and more than three months, respectively. It is revealed predominantly in the evolution of the CH area, but also very distinctly in its mean EUV intensity and the in-situ measured solar wind speed.

\item We find an influence of the differential rotation of the Sun on the CH, as seen by the change in orientation of the CHs main axis that corresponds with different rotation rates of different CH parts, higher at the equator and lower at higher latitudes. 
\item The correlation between CH area and peak solar wind speed of the associated HSS is strongest during the maximum phase of the CH. Weaker relations are found for the growing and decaying phase. This may be related to the 3 dimensional extension and propagation of the HSS, that is only measured in one specific location. Density, pressure and magnetic field parameter of the solar wind show no distinct relation to the CH area but we find a general evolutionary trend.

\end{enumerate}

We conclude that the CH shows an evolutionary pattern with different behaviors during the three phases. During the formation of the CH until it reaches its maximum, the parameters are found to be more stable compared to the decaying process. These different physical processes most likely are related to the underlying magnetic field and affect not only the surface properties but also the outflowing plasma in 1 AU distance. Part~II of this study investigates in detail the CH underlying photospheric magnetic field from which also a distinct 3-phase evolution is revealed \citep{2018heinemann_paperII}.

\section*{Acknowledgements}
The SDO and STEREO image data and the WIND, ACE and STEREO in-situ data is available by courtesy of NASA and the respective science teams. We acknowledge the support by the FFG/ASAP Program under grant no. 859729 (SWAMI). A.M.V.\,and M.T.\,acknowledge the Fonds zur F\"orderung wissenschaftlicher Forschung (FWF): P24092-N16 and V195-N16. S.J.H.\,acknowledges support from the JungforscherInnenfonds der Steierm\"arkischen Sparkassen.
\appendix



\startlongtable
\begin{deluxetable}{l c |c c c c c}
\tablecaption{Correlation Coefficients Overview\label{tab:corr}}
\tablehead{
\colhead{} & \colhead{}& \multicolumn{5}{c}{Pearson Correlation Coefficient}  \\
\colhead{Relation} &\colhead{Figure Nr.}& \colhead{$\mu_{\mathrm{P}}$} & \colhead{$\sigma_{\mathrm{P}}$} & \colhead{CI $90\%$} &\colhead{CI $95\%$} &\colhead{CI $99\%$}
} 
\startdata
$ I_{\mathrm{CH}} $ vs. $A_{\mathrm{CH}}$ 	& \ref{fig:int}(b) & $-0.57$ & $0.01$ & $[-0.56,-0.58]$ & $[-0.56,-0.59]$ & $[-0.55,-0.59]$   \\
$v_{\mathrm{SW}}$ vs. $A_{\mathrm{SW}}$ &\ref{fig:insitu4}(b)  &  $0.77$ & $0.08$ & $[0.61,0.87]$ & $[0.57,0.89]$ &$[0.46,0.91]$   \\ \hline \hline
\multicolumn{2}{ c }{ } & \multicolumn{5}{ c }{Spearman Correlation Coefficient}\\
\multicolumn{1}{ c }{Relation}&\multicolumn{1}{ c }{Figure Nr. } & $\mu_{\mathrm{S}}$ &$\sigma_{\mathrm{S}}$ & CI $90\%$ &CI $95\%$ &CI $99\%$ \\ \hline
$I_{\mathrm{CH}}$ vs. $A_{\mathrm{CH}}$ &\ref{fig:int}(b)  &$-0.60$ & $0.01$ & $[-0.58,-0.62]$ & $[-0.58,-0.62]$ & $[-0.57,-0.62]$\\
$v_{\mathrm{SW}}$ vs. $A_{\mathrm{SW}}$ &\ref{fig:insitu4}(b)  &  $0.72$ & $0.12$ & $[0.49,0.88]$ & $[0.43,0.89]$ & $[0.30,0.92]$   \\ \hline
\enddata
\end{deluxetable}





\bibliographystyle{aasjournal}

\end{document}